\documentclass[5p,times,twocolumn]{elsarticle}
\usepackage{amsmath,amssymb}
\usepackage{graphicx}
\usepackage{microtype}
\usepackage{xcolor}
\usepackage{booktabs}
\usepackage[colorlinks=true,linkcolor=blue!50!black,citecolor=blue!50!black,urlcolor=blue!50!black]{hyperref}
\graphicspath{{figures/}}

\journal{Physica A: Statistical Mechanics and its Applications}
\biboptions{sort&compress}
\newcommand{\rr}{\mathbf{r}}
\newcommand{\dd}{\mathrm{d}}

\begin{document}
\begin{frontmatter}
\title{Radicalization Kinetics under Algorithmic Exposure in a Stochastic Multiplex Model of Opinion Dynamics}
\author{Ruben E. Ara\'ujo\corref{cor1}}
\ead{rubenesteche@hotmail.com}
\cortext[cor1]{Corresponding author.}
\address{Independent Researcher}
\begin{abstract}
We study how physical mobility, algorithmic exposure, and repulsive social
influence interact in a stochastic multiplex model of opinion dynamics.
Agents diffuse in physical space while a directed digital network rewires
under a conserved attention budget, so digital exposure displaces rather
than supplements local interaction. With purely assimilative
bounded-confidence influence, opinion-blind long-range exposure reduces
locality-induced fragmentation whereas homophilic recommendation preserves
echo chambers. When a contested repulsive response to sufficiently distant
opinions is activated, this ordering reverses at the reference parameters:
a neutral platform reaches the maximal polarization permitted by the
bounded opinion space, controversy-seeking curation drives faster initial
separation but slows sharply near the boundary, and homophilic curation
delays radicalization by suppressing cross-bloc exposure. In a late-stage
symmetric two-bloc reduction, any curation kernel maps to a state-dependent
cross-bloc exposure profile $p(y)$ and an exact quadrature for the
radicalization time. Pointwise-ordered profiles inherit a global kinetic
ordering; crossing profiles yield target- and horizon-dependent rankings.
For similarity-driven curation the quadrature has a closed form involving
the exponential integral. Simulations, finite-size scans to $N=1600$,
structural controls, and a well-mixed particle comparison support the
mechanism. Heavy-tailed influence strengths are not required for the
inversion; in the well-mixed heavy-tail regime they additionally produce a
non-self-averaging stable-weighted asymptotic description. Finally,
opinion-independent Brownian mobility produces no detectable geographic
opinion structure in the explored regime, whereas opinion-dependent drift
produces spatial domains through a P\'eclet-controlled crossover near
$\chi\ell/D\sim1$.
\end{abstract}
\end{frontmatter}

\section{Introduction}
\label{sec:intro}

The most common mechanistic story about platforms and polarization runs
through similarity: recommendation algorithms feed users content they
already agree with, creating echo chambers whose members drift apart from
the rest of society \citep{sirbu2019, santos2021, baumann2020}. The story
suggests an obvious remedy---expose people to the other side. Yet the
best-known field experiment on this intervention found the opposite of the
intended effect: Republicans exposed for a month to a stream of liberal
content became substantially \emph{more} conservative \citep{bail2018}
(the broader field record is mixed: other interventions have reduced
polarization or left it unchanged \citep{levy2021, nyhan2023}). Social
psychology offers a candidate mechanism for such backfire---contested, as
we discuss below: influence is not always assimilative, and views far
beyond an agent's latitude of acceptance can repel rather than attract
\citep{jager2005, flache2017, tang2025}. That algorithmic mediation and
cross-cutting exposure can either polarize or depolarize is by now well
established \citep{starnini2025}: attraction--repulsion models show that
exposure can polarize intolerant populations \citep{axelrod2021,
sabinmiller2020}; polarization can rise with increasing connectivity
itself, without any suppression of interaction \citep{pham2026};
post-distribution and recommendation algorithms can
promote or mitigate polarization depending on their strategy
\citep{dearruda2022, pansanella2022, cinus2022}; mild random
``nudges'' toward out-group content
depolarize an echo-chambered population while excessive nudging
radicalizes it \citep{pal2023}; and collaborative-filtering
recommendations produce analytically tractable polarization phases
\citep{bellina2023}. What
these works establish qualitatively---that the sign of an exposure
intervention is not fixed---they do not provide a kinetic reduction that
compares curation strategies on a common quantitative basis. That is the
gap addressed here. Under a conserved attention budget, the late-stage
two-bloc dynamics maps a platform's curation kernel to a state-dependent
fraction of cross-bloc exposure in the repulsive zone. This exposure
profile controls the outward drift and, through an exact quadrature,
sets the finite-time radicalization scale---by which we mean, throughout,
growth of the mean absolute opinion $\langle|x|\rangle$ toward the
boundary of opinion space. Pointwise-ordered exposure profiles inherit a
global kinetic ordering; when profiles cross, their comparison is instead
target- and horizon-dependent and is determined by the full quadrature.

This paper builds a minimal model in which that question has a sharp
answer. Agents carry continuous opinions and interact through two layers
simultaneously: a physical layer, where encounters are structured by
geography---agents diffuse through a two-dimensional world by Brownian
motion and influence each other through a short-range bounded-confidence
kernel \citep{deffuant2000, hegselmann2002, alraddadi2024}---and a digital
layer, a directed attention graph whose links are continuously rewired by a
platform according to an \emph{engagement kernel} $E(\Delta)$: the
probability that content at opinion distance $\Delta$ is shown. A finite
attention budget couples the layers: the digital attention share $\lambda$
displaces physical interaction rather than adding to it. On top of this
multiplex substrate \citep{antonopoulos2018} we add the two
psychologically motivated ingredients: an
assimilation--indifference--repulsion influence function
\citep{jager2005, kurmyshev2011, tang2025} and heavy-tailed influence
strengths.

Our headline result is an inversion in the finite-time platform ranking.
With purely assimilative influence, the model reproduces the standard
picture---an opinion-blind platform acts like infinite-range mobility and
heals fragmentation, while algorithmic homophily freezes fragmentation
into spatially delocalized echo chambers. With the repulsive channel active
at the reference attention share, all three designs develop outward
radicalizing drift, but with sharply different finite-horizon outcomes: the
neutral platform reaches the maximal polarization permitted by the bounded
opinion space, controversy-seeking curation approaches a strong plateau,
and strong algorithmic homophily produces the slowest separation because
it suppresses cross-bloc encounters. The distinction is kinetic rather
than a universal asymptotic ranking: a controversy kernel can generate the
fastest initial separation while becoming slower than the other designs
near the boundary as its exposure profile changes with opinion distance.

The inversion can be understood directly from the structure of the
model, which we reduce analytically in three steps. In the late-stage
symmetric two-bloc reduction, the stationary fraction of attention slots
pointing across blocs,
$p(y) = E(2y) / [E(0) + E(2y)]$ for blocs at $\pm y$, controls the
radicalization rate $\dot y = 2\alpha_{\mathrm{tot}}\lambda\eta\, p(y)\,
y$. The resulting exposure profiles explain the distinct kinetics of the
three kernels: the neutral profile is constant, similarity-driven exposure
is increasingly suppressed as the blocs separate, and the controversy
profile crosses the neutral baseline, producing very rapid early
separation followed by strong finite-horizon slowing near the boundary. Within this deterministic
reduction, radicalization has \emph{no} threshold in the attention
share---any $\lambda > 0$ with nonzero cross-bloc exposure produces
outward drift---and the reduction yields, with no fitted parameters, an
exact finite-horizon crossover by quadrature together with a compact
local-rate estimate $\lambda_c^{(0)}(T)$ whose values span two orders of
magnitude across designs, are consistent with the simulated onsets of the
neutral and controversy-driven platforms, and predict that no
boundary-reaching crossover exists in the physical range for the
similarity-driven platform. And integrating the fast-rewiring,
well-mixed counterpart of the model reproduces the neutral
$\mathrm{Var}(x)$-versus-$\lambda$ curve of the spatial agent-based model
nearly quantitatively away from the steep onset region, and tracks the
controversy curve from below with a gap that narrows as $\lambda$ grows.

Two secondary results delimit the role of geography. First, in a phase
diagram spanning four decades of mobility $D$ and the full attention range
$\lambda$, the radicalized region at high $\lambda$ is independent of $D$:
the dependence of the final state on mobility becomes weak once digital
attention dominates the interaction budget. Second, under opinion-independent Brownian mobility we detect
no geographic opinion structure at \emph{any} mobility---a null result
whose first-order part follows from translation invariance and that
qualifies a common intuition about local echo chambers. Within this model, geographic opinion structure requires
opinion--position coupling: adding a Schelling-type homophilic drift of
strength $\chi$ \citep{pasimeni2025, djurdjevac2024}, we find that spatial
opinion domains emerge through a P\'eclet-controlled crossover centered near
$\chi \ell / D \sim 1$, where homophilic drift and diffusion become
comparable over the interaction range $\ell$.

The ingredients of the model each have a literature, recently
systematised by \citet{starnini2025}: bounded confidence
\citep{deffuant2000, hegselmann2002, lorenz2007, castellano2009,
bernardo2024}, mobile agents \citep{alraddadi2024}, coevolving networks
\citep{holme2006}, multiplex opinion formation \citep{antonopoulos2018},
algorithmic bias and link recommendation \citep{sirbu2019, santos2021,
pansanella2022}, activity-driven echo chambers \citep{baumann2020},
stubborn agents \citep{tian2018}, adaptive confidence \citep{li2025}, and
coupled opinion--position dynamics \citep{djurdjevac2024, pasimeni2025}.
Closest in spirit is the post-distribution model of \citet{dearruda2022},
in which an algorithm's distribution strategy determines whether
polarization is promoted or mitigated under attractive and repulsive
interactions; \citet{bellina2023} derive stationary polarization phases
for collaborative-filtering recommendations, whereas our object is the
\emph{kinetics} of attraction--repulsion dynamics under conserved
attention. The contribution here relative to that line is threefold:
(i) a conserved-attention multiplex construction in which digital exposure
displaces physical interaction rather than simply adding another channel;
(ii) within the late-stage two-bloc reduction, a mapping from an arbitrary
curation kernel to a state-dependent repulsive-exposure profile whose
quadrature yields explicit finite-time radicalization scales; and (iii) a
quantitative comparison principle for platform designs---direct when their
exposure profiles are pointwise ordered and target- and horizon-dependent
when the profiles cross---tested against the full spatial model and a
well-mixed particle counterpart without fitting the onset data.

\section{Model}
\label{sec:model}

\subsection{State variables}

Each of $N$ agents carries a position $\rr_i(t) \in [0,L)^2$ with periodic
boundary conditions and an opinion $x_i(t) \in [-1,1]$. Agent $i$ may also
carry an influence strength $s_i$ and a stubbornness flag. The digital
layer is a directed graph $A_{ij}(t) \in \{0,1\}$, where $A_{ij}=1$ means
that the platform currently shows content by agent $j$ to agent $i$. Each
agent has a fixed number $k$ of \emph{attention slots}, $\sum_j A_{ij} = k$
for all $i$ and $t$: attention is conserved, and new digital connections
can only be acquired by dropping old ones.

\subsection{Dynamics}

Positions follow
\begin{equation}
  \dd\rr_i = \chi\, \mathbf{f}_i\,\dd t + \sqrt{2D}\,\dd\mathbf{W}_i(t),
  \label{eq:motion}
\end{equation}
where $D$ is the diffusion coefficient and $\mathbf{f}_i$ is an optional
homophilic drift (Level 5 below; $\chi = 0$ elsewhere): the
kernel-weighted mean of minimum-image unit vectors pointing toward
compatible neighbours ($|x_j-x_i|<\epsilon$) and away from incompatible
ones within the same spatial cutoff used below. The signed directional
contributions are normalized by their total kernel weight, and the drift
is set to zero when no contributing neighbour is present. Opinions obey
the It\^o equation
\begin{align}
  \dd x_i =\;&
  \alpha_i \,
  \frac{\sum_j K(r_{ij})\, B_\epsilon(x_j - x_i)\,(x_j - x_i)}
       {\sum_j K(r_{ij})\, B_\epsilon(x_j - x_i)} \,\dd t
  \nonumber\\
  &+
  \beta_i \,
  \frac{\sum_j A_{ij}\, s_j\, F(x_i, x_j)}
       {\sum_j A_{ij}\, s_j} \,\dd t
  \;+\;
  \sigma_x\, \dd B_i(t),
  \label{eq:opinion}
\end{align}
where Euler--Maruyama updates are projected onto $[-1,1]$ after each step
(a reflecting-boundary control appears in Sec.~\ref{sec:robustness}),
$r_{ij}$ is the
minimum-image distance between agents, sums run over $j \ne i$, and the
physical drift is set to zero for agents with no compatible neighbour
inside the kernel cutoff---a common event in the sub-percolation regime
studied below. The physical kernel is a truncated Gaussian,
\begin{equation}
  K(r) = e^{-r^2 / 2\ell^2}\, \mathbf{1}(r < 3\ell),
  \label{eq:kernel}
\end{equation}
and $B_\epsilon(\Delta) = \mathbf{1}(|\Delta| < \epsilon)$ implements
bounded confidence. The truncation is not cosmetic: because the drift is
normalised, an untruncated Gaussian would let an agent with no nearby
compatible peers average with the entire population at $O(1)$ rate through
the exponential tail, silently removing the mobility scale from the
problem.

The influence function has the assimilation--indifference--repulsion form
\begin{equation}
  F(x_i, x_j) =
  \begin{cases}
    x_j - x_i, & |x_j - x_i| < \epsilon_1, \\
    0, & \epsilon_1 \le |x_j - x_i| < \epsilon_2, \\
    -\eta\,(x_j - x_i), & |x_j - x_i| \ge \epsilon_2,
  \end{cases}
  \label{eq:influence}
\end{equation}
with the repulsive branch ($\eta > 0$) active at Level 4; at Levels
2--3, $F$ reduces to bounded confidence with threshold
$\epsilon_1 = \epsilon$. This three-zone form descends from
social-judgment-theory models \citep{jager2005, kurmyshev2011}. The
empirical evidence for the repulsive branch is mixed: stochastic
actor-oriented analyses of longitudinal network data support it
\citep{tang2025}, while controlled laboratory experiments find little
evidence of negative shifts \citep{takacs2016}; Level 4 should be read as
exploring the consequences of this contested mechanism, not as assuming it
settled.

\subsection{Attention budget}

The rates in Eq.~\eqref{eq:opinion} satisfy
\begin{equation}
  \alpha_i = \alpha_{\mathrm{tot}}\,(1 - \lambda),
  \quad
  \beta_i = \alpha_{\mathrm{tot}}\,\lambda,
  \quad \lambda \in [0,1],
  \label{eq:attention}
\end{equation}
so the digital attention share $\lambda$ measures how much of a finite
information-processing capacity is devoted to the platform. Increasing
digital exposure \emph{displaces} physical influence rather than adding to
it, alongside the hard slot constraint $\sum_j A_{ij} = k$.

\subsection{The platform: engagement-driven rewiring}

In each interval $\dd t$, agent $i$ replaces one uniformly chosen attention
slot with probability $\rho\,\dd t$; the platform selects the replacement
source $j$ with probability
\begin{equation}
  P(i \to j) \;\propto\; E\!\left(|x_i - x_j|\right)\, s_j,
  \label{eq:rewire}
\end{equation}
where $E$ is the platform's \emph{engagement kernel}; the replacement is
sampled among non-current sources, a finite-$k$ correction of order $k/N$
to the stationary exposure distribution used in Sec.~\ref{sec:theory},
vanishing as $N \to \infty$. We compare three designs:
\begin{equation}
  \begin{gathered}
  E_{\mathrm{sim}}(\Delta) = e^{-\gamma \Delta},
  \qquad
  E_{\mathrm{neu}}(\Delta) = 1,
  \\
  E_{\mathrm{con}}(\Delta) = e^{-(\Delta - \delta)^2 / 2 w^2}.
  \end{gathered}
  \label{eq:engagement}
\end{equation}
The similarity kernel models algorithmic homophily with strength $\gamma$
\citep{sirbu2019, santos2021}; the neutral kernel is an opinion-blind
baseline; the controversy kernel encodes a platform that maximises
engagement by promoting content at a preferred ideological distance
$\delta$, close in spirit to the post-distribution algorithms of
\citet{dearruda2022}.

\subsection{Model levels, heterogeneity, and observables}

The model is built in five nested levels: (1) Brownian motion + bounded
confidence; (2) static opinion-neutral digital links; (3) adaptive
similarity-driven rewiring; (4) general engagement kernels + repulsion +
heavy-tailed influence strengths (optional stubborn agents
\citep{tian2018}); (5) homophilic mobility $\chi > 0$ added to the purely
physical model of Level 1. Influence
strengths are drawn as $s_i = 1 + z_i$ with $z_i$
Lomax-distributed---density tail $p(s) \sim s^{-\kappa}$, i.e.\ survival
exponent $\kappa - 1$---and normalised to unit mean per realisation. At
the reference $\kappa = 2.5$ the variance of $s$ is infinite: a
deliberate ``influencer'' feature whose finite-$N$ fluctuations play a
visible role in Sec.~\ref{sec:results-threshold}.

We measure: polarization $P = \operatorname{Var}(x)$; extremism
$\langle |x| \rangle$; the number of opinion clusters $n_c$ (gaps larger
than $0.05$ in sorted opinion space, groups of at least two agents); a
categorical state (consensus / polarization / fragmentation for
$n_c \le 1$ / $= 2$ / $\ge 3$; states with
$\operatorname{Var}(x) < 10^{-2}$ are additionally classed as consensus,
an override that never fires in the reported data; note the scalar $P$
and the two-cluster categorical state share the word ``polarization'');
spatial autocorrelation of opinions via
Moran's $I$ (with Gaussian spatial weights of scale $\ell$) and via the
\emph{local agreement gap}
$\Pr(|x_i - x_j| < \epsilon \mid r_{ij} < 3\ell) - \Pr(|x_i - x_j| <
\epsilon)$; opinion assortativity $\rho$ of the digital graph; modularity
$Q$ of the digital graph with respect to opinion clusters; and mean local
disagreement $\langle |x_i - \bar{x}_{\partial i}| \rangle$ between an
agent and the average opinion it is shown.

\subsection{Numerical integration}

Equations \eqref{eq:motion}--\eqref{eq:opinion} are integrated by
Euler--Maruyama. Unless stated otherwise, $N = 200$, $L = 1$,
$\dd t = 0.02$, $\alpha_{\mathrm{tot}} = 1$, $\epsilon = \epsilon_1 = 0.3$,
$\epsilon_2 = 0.9$, $\eta = 0.4$, $\ell = 0.02$, $k = 10$, $\rho = 5$,
$\gamma = 4$, $\delta = 0.8$, $w = 0.2$, and $\kappa = 2.5$. Opinion noise
is $\sigma_x = 0.02$ whenever the digital layer is adaptive (Levels 3--4,
including the phase diagram) and $\sigma_x = 0$ in the purely physical and
static-layer experiments (Levels 1, 2, and 5). Horizons are $T = 60$
(Levels 1--3 and 5), $T = 80$ (Level 4, the robustness scans, and the
attention sweeps), and $T = 50$ (the phase diagram). In system-size
scans the box is rescaled as $L = \sqrt{N/200}$ so that density, kernel
range and digital degree stay constant. With $\ell = 0.02$ the mean
physical contact degree is $N\pi(3\ell)^2/L^2 \approx 2.3$, below the
percolation threshold of the random geometric graph, so the instantaneous
contact network is spatially fragmented and locality is meaningful.
Statistics use $6$--$12$ realisations per point ($3$ for the phase
diagram); error bars denote the standard deviation across realisations.
Halving $\dd t$ twice shifts the Level-4 polarization values by amounts
comparable to that spread ($\lesssim 0.05$, concentrated in the
similarity platform and directed upward---i.e.\ against the inversion,
which is therefore conservative) without affecting the platform
ordering.

\section{Theory: two-bloc reduction and well-mixed counterpart}
\label{sec:theory}

\subsection{Late stage: cross-bloc attention controls radicalization}
\label{sec:twobloc}

Consider the late-stage configuration observed at Level 4: two equal blocs
at opinions $\pm y$. When attention slots renew quickly compared with the
bloc drift (slot renewal rate $\rho/k$ against the rate
$2\alpha_{\mathrm{tot}}\lambda\eta p$ of Eq.~\eqref{eq:bloc} below---a
condition that is only marginally met for the neutral kernel at our
reference parameters, so the agreement found below also attests to the
reduction's robustness against partial slot memory), the stationary
probability that an attention slot of a $+$-bloc agent
points into the $-$ bloc follows from Eq.~\eqref{eq:rewire}:
\begin{equation}
  p(y) \;=\; \frac{E(2y)}{E(0) + E(2y)}.
  \label{eq:p}
\end{equation}
(Equation~\eqref{eq:p} is exact in the mean for homogeneous strengths
$s_i = 1$, which is therefore the natural theoretical baseline; a
homogeneous-strength control reproduces the inversion unchanged
(Sec.~\ref{sec:robustness}). With heavy-tailed strengths, slots are both
sampled and weighted proportionally to $s$, and the realised cross-bloc
exposure fluctuates around the homogeneous-strength prediction: for a
realised population the slot-sampling probability is
$E(2y) S_{\mp} / [E(0) S_{\pm} + E(2y) S_{\mp}]$, with $S_{\pm}$ the two
blocs' realised strength totals, which reduces to Eq.~\eqref{eq:p} as
their ratio approaches unity---guaranteed by the finite mean of the
strength law, though convergence with $N$ is slow for our
infinite-variance reference distribution. Given the slot types, the
expected $s$-weighted drift then equals the unweighted one by slot
exchangeability, since strengths are independent of bloc membership; see
Sec.~\ref{sec:meanfield}.) Within-bloc content
exerts no drift ($F = 0$ at $\Delta = 0$), and
cross-bloc content is repulsive once $2y \ge \epsilon_2$, so the bloc
positions obey
\begin{equation}
  \frac{\dd y}{\dd t} \;=\; 2\,\alpha_{\mathrm{tot}}\,\lambda\,\eta\;
  p(y)\; y,
  \qquad 2y \ge \epsilon_2,
  \label{eq:bloc}
\end{equation}
with no physical restoring force: once the blocs are separated beyond the
confidence bound, the physical layer only pulls agents toward their own
bloc's mean. Equation~\eqref{eq:bloc} is the central statement of this
paper within the late-stage symmetric two-bloc reduction: any curation
kernel $E$ is mapped to a state-dependent cross-bloc exposure profile
$p(y)$, whose quadrature sets the radicalization time
[Eq.~\eqref{eq:tradgen} below]. If two profiles are pointwise ordered over
the traversed opinion range, their radicalization times inherit the
opposite ordering directly. If the profiles cross, no single global
ranking exists independently of the target and observation horizon; the
full quadrature is then required. The three kernels of
Eq.~\eqref{eq:engagement} illustrate both cases: neutral and similarity
are pointwise ordered, whereas the controversy profile crosses the
neutral baseline as the blocs separate. Their distinct $p(y)$ profiles
therefore produce the following kinetics:

\begin{itemize}
  \item \textbf{Neutral:} $p \equiv 1/2$. The drift
  $\alpha_{\mathrm{tot}}\lambda \eta y$ never
  shuts off; opinions are driven to the boundary and pinned at $\pm 1$.
  \item \textbf{Controversy-driven:} $p(y) \approx 1$ for
  $2y \gtrsim \epsilon_2$ (cross-bloc content is exactly what engages), so
  separation is fast. For the Gaussian kernel of Eq.~\eqref{eq:engagement},
  $p_{\mathrm{con}}(y)=1/2$ at $y=\delta$ (besides the trivial $y=0$
  solution), so the controversy profile crosses the neutral baseline there;
  for $y>\delta$ its cross-bloc exposure is smaller than neutral. As
  $2y \to 2$, both $E(2y)$ and $E(0)$ are far from the engagement peak and
  $p \to E(2)/[E(0)+E(2)] \approx e^{2(\delta-1)/w^2}$, which for
  $\delta<1$ is exponentially small. The dynamics therefore becomes
  exponentially slow near the boundary, producing an
  \emph{apparent finite-horizon plateau}: the platform continues to exert a
  nonzero outward drift, but its cross-bloc exposure becomes so small that
  further motion is negligible on the simulated horizons.
  \item \textbf{Similarity-driven:} $p(y) = e^{-2\gamma y}/(1 +
  e^{-2\gamma y})$ is small for $\gamma y \gtrsim 1$. Radicalization
  proceeds, but at a rate suppressed by $e^{-2\gamma y}$: a slow outward
  creep fed by the few cross-cutting links that stochastic rewiring keeps
  producing.
\end{itemize}

These three behaviours are exactly what the simulations show
(Sec.~\ref{sec:level4}): pinning at $\pm 1$ for neutral
($\langle|x|\rangle \approx 1.00$), an apparent plateau short of the boundary
for controversy ($\approx 0.95$), and slow partial radicalization for
similarity ($\approx 0.79$ at $T = 80$). Note also that within the
deterministic reduction Eq.~\eqref{eq:bloc} has no positive critical
$\lambda$: any $\lambda > 0$ with nonzero cross-bloc exposure produces
outward drift. What varies across designs---by orders
of magnitude---is the rate, which we quantify next.

\subsection{Onset: radicalization is rate-limited, not threshold-limited}
\label{sec:threshold}

\begin{figure*}[!tp]
  \centering
  \includegraphics[width=0.88\linewidth]{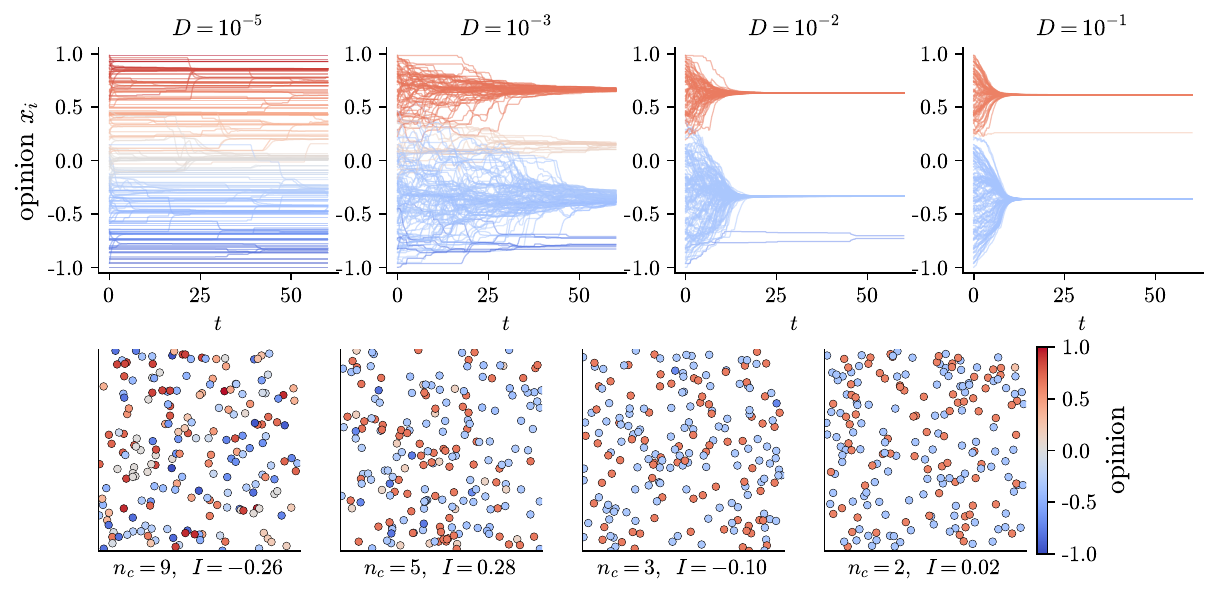}
  \caption{Purely physical dynamics (Level 1). Top: opinion trajectories
  $x_i(t)$ for increasing diffusion coefficient $D$; colours encode final
  opinions. Bottom: final spatial configurations. Panel captions report
  the cluster count $n_c$ and Moran's $I$ of the single realisation shown;
  individual-run values of $I$ fluctuate around a zero ensemble mean
  (s.d.\ $\approx 0.13$ across seeds, single-run extremes up to
  $|I| \approx 0.35$; Fig.~\ref{fig:mobilitysweep}). Low mobility freezes many coexisting
  clusters; high mobility recovers the mean-field bounded-confidence
  outcome. Note the absence of spatial colour domains at any $D$.}
  \label{fig:mobility}
\end{figure*}

\begin{figure}[!tbp]
  \centering
  \includegraphics[width=\linewidth]{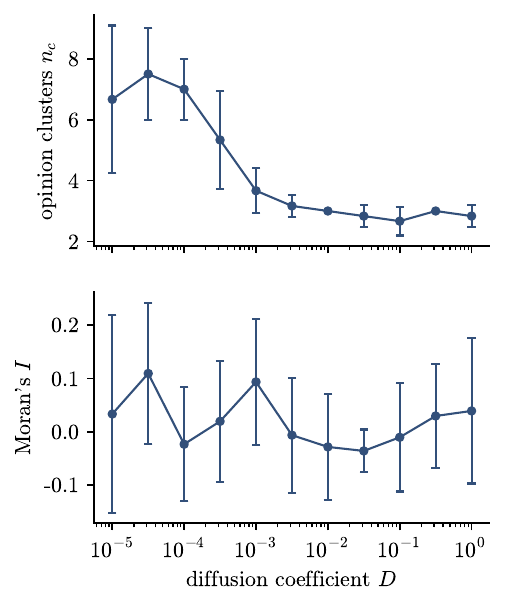}
  \caption{Level-1 sweep over the diffusion coefficient. Top: mean number
  of surviving opinion clusters. Bottom: spatial autocorrelation of
  opinions measured by Moran's $I$ (mean $\pm$ s.d.\ over 6
  realisations). Mobility interpolates between local fragmentation and
  mean-field behaviour while spatial opinion autocorrelation stays at
  noise level throughout.}
  \label{fig:mobilitysweep}
\end{figure}

\begin{figure*}[!tp]
  \centering
  \includegraphics[width=\linewidth]{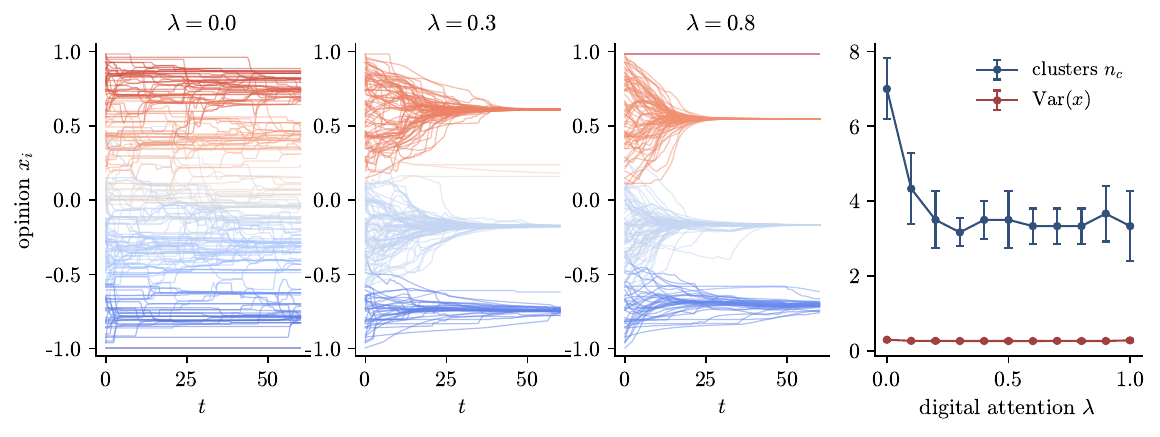}
  \caption{Level 2: static random digital layer at low mobility
  ($D = 10^{-4}$). Left three panels: opinion trajectories for increasing
  digital attention share $\lambda$. Right: cluster count and polarization
  versus $\lambda$ (mean $\pm$ s.d.).}
  \label{fig:static}
\end{figure*}

One might expect a critical attention share below which the physical
layer's assimilative pull protects the population. A deliberately
na\"ive stability estimate around the initially uniform state gives such
thresholds: for an edge agent, the bounded-confidence physical term has
inward magnitude $(1-\lambda)\epsilon/2$, while the digital term is the
engagement-weighted repulsive drift generated by opinions beyond
$\epsilon_2$; equating these two contributions at the reference
parameters gives $\lambda^*\approx0.34$ for the neutral kernel and
$0.55$ for the controversy kernel. These values are used only as a
foil---the simulations show radicalizing drift already at
$\lambda\approx0.05$ (Sec.~\ref{sec:results-threshold}). The failure is
mechanistic rather than numerical: bounded-confidence fragmentation moves
the system away from the uniform state before that putative instability
can control the late-time dynamics.
For attention shares bounded away from $\lambda=1$, bounded-confidence
fragmentation forms first on the timescale $\sim(1-\lambda)^{-1}$,
producing blocs near $\pm y_0$ with $y_0 \approx 0.6$ (the
bounded-confidence cluster positions at $\epsilon=0.3$, read off from
Figs.~\ref{fig:mobility} and \ref{fig:static}, not fitted to the
threshold data). Since $2y_0>\epsilon_2$, this fragmented state is
unstable to repulsion for any $\lambda>0$: once separated beyond the
confidence bound, the physical layer exerts no restoring force between
blocs (Sec.~\ref{sec:twobloc}). The timescale separation necessarily
weakens as $\lambda\to1$, where the physical layer vanishes; in that
limit $y_0$ should be interpreted as a late-stage initialization of the
two-bloc reduction rather than as the outcome of a parametrically faster
physical transient.

What remains is a rate. Equation~\eqref{eq:bloc} is separable, so the
time to traverse from $y_0$ to $y_f$ follows exactly by quadrature,
\begin{equation}
  t(y_0 \to y_f)
  \;=\;
  \frac{1}{2\,\alpha_{\mathrm{tot}}\,\lambda\,\eta}
  \int_{y_0}^{y_f} \frac{\dd y}{y\, p(y)},
  \label{eq:tradgen}
\end{equation}
which maps any curation kernel to a radicalization timescale through its
exposure profile alone. Because $t \propto 1/\lambda$, the \emph{exact}
finite-horizon crossover for reaching a target $y_f$ within an observation
window $T$ follows immediately:
\begin{equation}
  \lambda_c(T; y_f)
  \;=\;
  \frac{1}{2\,\alpha_{\mathrm{tot}}\,\eta\, T}
  \int_{y_0}^{y_f} \frac{\dd y}{y\, p(y)},
  \label{eq:lambdacexact}
\end{equation}
whenever the right-hand side does not exceed one; otherwise the target is
unreachable within the window at any attention share. For the similarity
kernel, the quadrature is available in closed form because
$p_{\mathrm{sim}}(y)=1/[1+\exp(2\gamma y)]$:
\begin{equation}
  t_{\mathrm{sim}}(y_0\to y_f)
  = \frac{\ln(y_f/y_0)
  + \operatorname{Ei}(2\gamma y_f)
  - \operatorname{Ei}(2\gamma y_0)}
  {2\,\alpha_{\mathrm{tot}}\,\lambda\,\eta},
  \label{eq:tradsim}
\end{equation}
where $\operatorname{Ei}$ is the exponential integral. At the reference
parameters with $y_0=0.6$ and $y_f=1$, Eq.~\eqref{eq:tradsim} gives
$t_{\mathrm{sim}}\simeq507.7/\lambda$, so
Eq.~\eqref{eq:lambdacexact} gives $\lambda_c = 6.35$: the boundary cannot
be reached within $T=80$ even at $\lambda=1$, and no boundary-reaching
crossover exists in the physical interval $[0,1]$. For the controversy
kernel the quadrature is evaluated numerically: from $y_0 = 0.6$,
$t(0.6 \to 0.90) \simeq 3.2/\lambda$,
$t(0.6 \to 0.95) \simeq 32.1/\lambda$, and
$t(0.6 \to 1) \simeq 482/\lambda$. Fast initial radicalization is followed
by a dramatic slowing: the plateau level $\langle|x|\rangle \approx 0.95$
observed in the simulations is reached within $T = 80$ only for
$\lambda \gtrsim \lambda_c(80; 0.95) \approx 0.40$, while the boundary itself lies far beyond the principal $T=80$
observation window and is not reached on the simulated horizons.

For a compact local-rate estimate applicable to all three designs, we
also use the constant-exposure approximation
$p(y) \approx p_0 = p(y_0)$, which gives the \emph{local} timescale
\begin{equation}
  t_{\mathrm{loc}}(\lambda)
  \;=\;
  \frac{\ln (y_f / y_0)}{2\,\alpha_{\mathrm{tot}}\,\lambda\,\eta\, p_0},
  \label{eq:trad}
\end{equation}
and the corresponding local-rate crossover estimate $t_{\mathrm{loc}} = T$:
\begin{equation}
  \lambda_c^{(0)}(T)
  \;=\;
  \frac{\ln (y_f / y_0)}{2\,\alpha_{\mathrm{tot}}\,\eta\, p_0\, T}.
  \label{eq:lambdac}
\end{equation}
The frozen-$p_0$ approximation is exact for the neutral kernel, whose
$p \equiv 1/2$ is state-independent, so there
$\lambda_c^{(0)}$ coincides with Eq.~\eqref{eq:lambdacexact}; for the
state-dependent kernels it characterizes the initial rate only.
Table~\ref{tab:rates} evaluates Eqs.~\eqref{eq:trad}--\eqref{eq:lambdac}
at the reference parameters with no fitting. The cross-bloc slot fraction
$p_0$ spans two orders of magnitude across the designs, and with it the
local rate: for the neutral platform
$\lambda_c^{(0)} = \lambda_c \approx 0.016$ and for the controversy
platform $\lambda_c^{(0)} \approx 0.008$---any noticeable digital
attention starts radicalization within the horizon, though for
controversy the exact plateau-reaching crossover is the larger
$\lambda_c(80;0.95) \approx 0.40$ above. For the similarity platform
$\lambda_c^{(0)} \approx 0.98$ is a local-rate figure only: the exact
quadrature shows that no boundary-reaching crossover exists in the
physical range, and even full digital attention only partially
radicalizes the population in the same window---exactly the smooth
partial growth observed in the simulations. ``Protection'' by homophilic
curation is therefore kinetic, not asymptotic. At the initial two-bloc
state its local rate is suppressed by the factor $p_0$ relative to maximal
cross-bloc exposure; relative to the neutral baseline $p=1/2$, the local
timescale is enhanced by $(2p_0)^{-1}$. Because $p_{\mathrm{sim}}(y)$
decreases further as the blocs separate, the exact quadrature gives an
even larger finite-time delay without changing the destination within the
deterministic reduction.

Two caveats delimit Table~\ref{tab:rates}. The constant-$p_0$
approximation is exact for the neutral kernel but conservative for the
similarity kernel, whose $p(y)$ decays further as the blocs separate: the
exact time $507.7/\lambda$ is more than sixfold longer than
$78.2/\lambda$, so the true kinetic protection is even stronger than the
tabulated estimate suggests. For the controversy kernel, $p(y)$ collapses
as $2y \to 2$ (Sec.~\ref{sec:twobloc}), so $t_{\mathrm{loc}}$ measures the
fast initial separation rate rather than the time to any fixed target;
target-crossing times follow from Eq.~\eqref{eq:tradgen}.

\begin{table}[!t]
  \caption{Two-bloc local-rate analysis, Eqs.~\eqref{eq:p} and
  \eqref{eq:trad}--\eqref{eq:lambdac}, at the reference parameters
  ($y_0 = 0.6$, $y_f = 1$, $\eta = 0.4$, $\gamma = 4$, $\delta = 0.8$,
  $w = 0.2$, $T = 80$, $\alpha_{\mathrm{tot}} = 1$). No parameters are
  fitted to these data; $y_0$ is the observed
  bounded-confidence cluster position. The tabulated $\lambda_c^{(0)}$
  are frozen-$p_0$ estimates, exact only for the neutral kernel; exact
  target-crossing crossovers follow from Eq.~\eqref{eq:lambdacexact} and
  are quoted in Sec.~\ref{sec:threshold}.}
  \label{tab:rates}
  \begin{tabular}{lccc}
    \toprule
    Platform & $p_0$ & $t_{\mathrm{loc}}(\lambda)$ & $\lambda_c^{(0)}(T{=}80)$ \\
    \midrule
    Similarity-driven & $0.0082$ & $78.2/\lambda$ & $0.98$ \\
    Neutral & $0.500$ & $1.28/\lambda$ & $0.016$ \\
    Controversy-driven & $0.998$ & $0.64/\lambda$ & $0.008$ \\
    \bottomrule
  \end{tabular}
\end{table}

\subsection{Well-mixed limit}
\label{sec:meanfield}

\begin{figure*}[!tp]
  \centering
  \includegraphics[width=\linewidth]{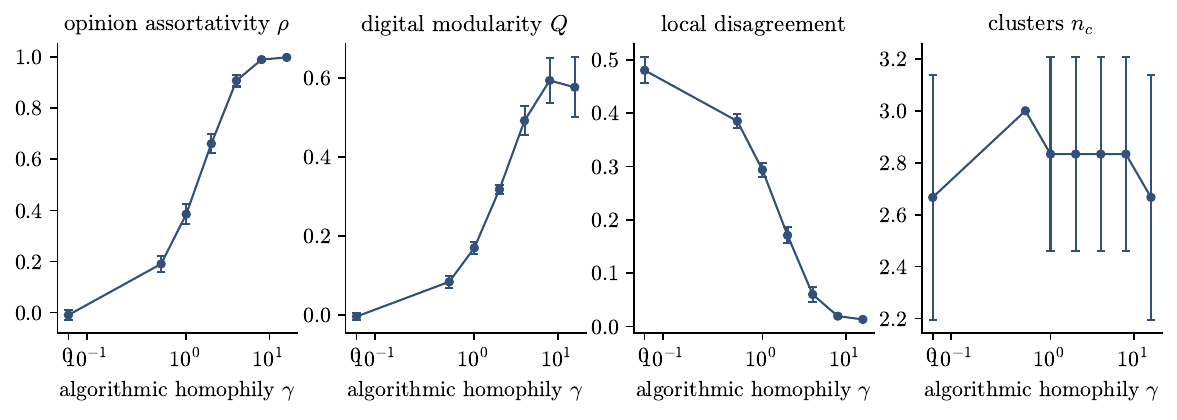}
  \caption{Level 3: adaptive digital layer with similarity-driven
  recommendation, $\lambda = 0.5$, $D = 10^{-3}$. From left to right:
  opinion assortativity of the digital graph, modularity with respect to
  opinion clusters, mean local disagreement, and cluster count as
  functions of the algorithmic homophily $\gamma$ (mean $\pm$ s.d.\ over 6
  runs).}
  \label{fig:adaptive}
\end{figure*}

\begin{figure}[!tbp]
  \centering
  \includegraphics[width=\linewidth]{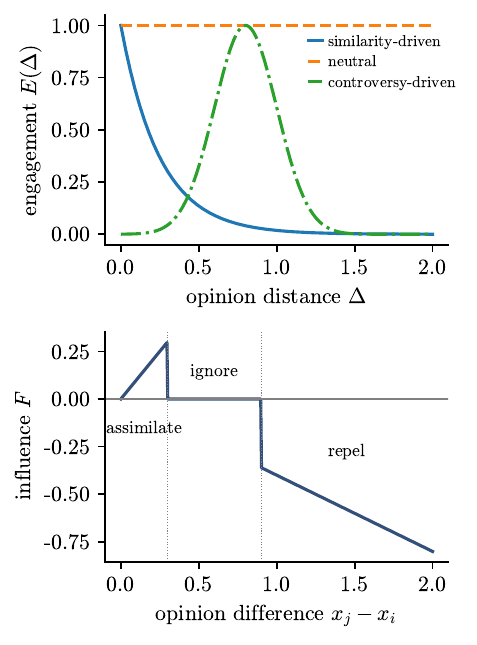}
  \caption{Level-4 ingredients. Top: the three engagement kernels of
  Eq.~\eqref{eq:engagement} ($\gamma = 4$, $\delta = 0.8$, $w = 0.2$).
  Bottom: the assimilation--indifference--repulsion influence function of
  Eq.~\eqref{eq:influence} ($\epsilon_1 = 0.3$, $\epsilon_2 = 0.9$,
  $\eta = 0.4$).}
  \label{fig:kernels}
\end{figure}

In the fast-rewiring, well-mixed counterpart of the model, each agent
feels the global bounded-confidence mean with weight $1 - \lambda$ and a
population-wide engagement-weighted influence average with weight
$\lambda$,
\begin{align}
  \dd x =\;&
  (1-\lambda)\,
  \frac{\mathbb{E}\big[(x' - x)\,\mathbf 1(|x'-x|<\epsilon)\big]}
       {\mathbb{P}\big(|x'-x|<\epsilon\big)}\, \dd t
  \nonumber\\
  &+ \lambda\,
  \frac{\sum_j s_j^{2}\,E(|x_j-x|)\, F(x, x_j)}
       {\sum_j s_j^{2}\,E(|x_j-x|)}\, \dd t
  + \sigma_x\, \dd B,
  \label{eq:mkv}
\end{align}
where $x'$ is an independent draw from the population law and the digital
sum runs over the whole population. The squared strengths arise because
slots are both \emph{sampled} and \emph{weighted} proportionally to $s$.
The status of Eq.~\eqref{eq:mkv} as $N \to \infty$ depends on the tail of
the strength law. For $\kappa > 3$, $\mathbb{E}[s^2]$ is finite, the
self-normalised digital drift self-averages, strengths drop out entirely
(they are independent of opinions), and Eq.~\eqref{eq:mkv} converges to a
deterministic McKean--Vlasov dynamics. At our reference $\kappa = 2.5$,
however, $\mathbb{E}[s^2] = \infty$ and the size-biased slot-strength law
has infinite mean, so the digital drift does \emph{not} self-average: it
remains a random variable dominated by the few largest strengths as
population size grows---influencer fluctuations therefore do not vanish
through conventional self-averaging in this reference heavy-tail regime. For the finite-mean regime $\kappa>2$, the slot-level two-bloc
expectation in Sec.~\ref{sec:twobloc} remains strength-free because slot
types are independent of the selected strengths. The population-wide
limit in Eq.~\eqref{eq:mkv} is different when $2<\kappa<3$. Writing
$Z=s^2$, its survival tail is regularly varying with index
$\alpha=(\kappa-1)/2\in(1/2,1)$. Under the standard marked
regular-variation approximation, treating the strength marks as
asymptotically independent of bloc membership, the normalized sums in
Eq.~\eqref{eq:mkv} converge in distribution to positive
$\alpha$-stable random measures rather than to their expectations
\citep{resnick2007}. Multiplying the regularly varying weight $Z$ by the
bounded positive engagement mark $E$ rescales its tail intensity by
$E^{\alpha}$, so the corresponding stable control measure is proportional
to $E^{\alpha}$. In this asymptotic description the well-mixed digital
drift remains random, while its annealed mark weighting is proportional to
$E^{\alpha}$ rather than $E$. In the two-bloc notation,
the corresponding annealed cross-bloc fraction is
\begin{equation}
  p_{\alpha}(y)=
  \frac{E(2y)^{\alpha}}
       {E(0)^{\alpha}+E(2y)^{\alpha}},
  \qquad \alpha=\frac{\kappa-1}{2},
  \label{eq:palpha}
\end{equation}
which reduces to the ordinary deterministic weighting only when the
second moment becomes finite. At the reference $\kappa=2.5$,
$\alpha=3/4$: heavy-tailed influencer fluctuations therefore flatten
algorithmic selectivity in the population-wide annealed limit while
remaining non-self-averaging realization by realization. This distinction
does not underwrite the inversion itself, which persists for $s_i=1$ and
$\kappa=4$ (Sec.~\ref{sec:robustness}); it clarifies instead why the
heavy-tailed well-mixed comparison is a stochastic limit rather than a
deterministic McKean--Vlasov closure.

We therefore compare the spatial agent-based model against a direct
integration of Eq.~\eqref{eq:mkv} as a spaceless well-mixed particle
model ($M = 600$, 12 realisations, same strength law), i.e.\ the model
stripped of space and of the finite attention window, in
Sec.~\ref{sec:results-threshold}: with no fitted parameters it reproduces
the neutral curve nearly quantitatively away from the steep onset region,
tracks the controversy-driven curve from below with a gap that narrows as
$\lambda$ grows, and tracks the shape of the similarity-driven
curve, which it systematically underestimates because the slot-level
sampling noise it discards (each spatial agent draws only $k = 10$
sources) is what feeds the similarity platform's slow creep. The
inversion is thus a property of the opinion dynamics and the engagement
weighting, not an artefact of space or slow rewiring.

\section{Results}
\label{sec:results}

\subsection{Baselines: mobility and neutral connectivity (Levels 1--2)}
\label{sec:baselines}

\begin{figure*}[!tp]
  \centering
  \includegraphics[width=\linewidth]{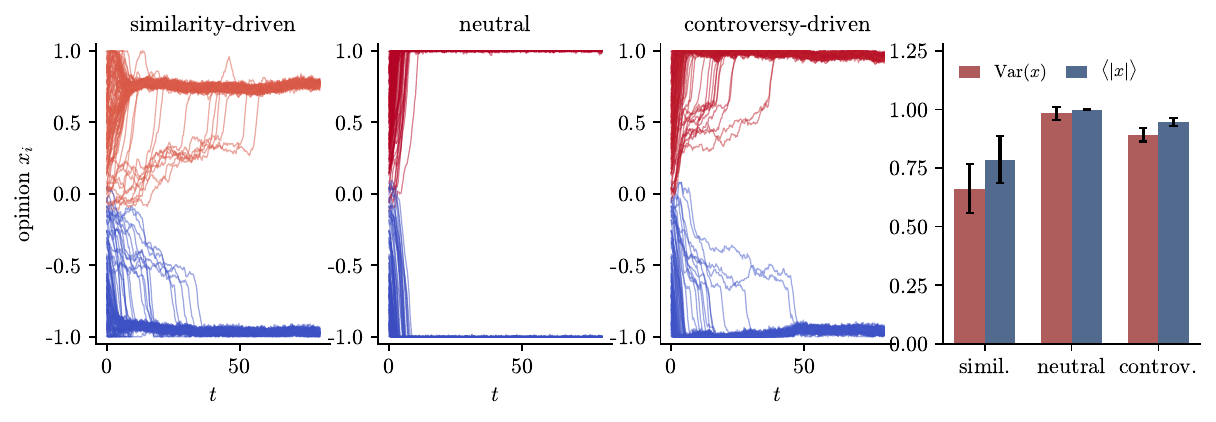}
  \caption{Level 4: platform designs compared under repulsive influence
  and heavy-tailed influence strengths ($\lambda = 0.6$). Left three
  panels: representative opinion trajectories. Right: polarization
  $\operatorname{Var}(x)$ and extremism $\langle |x| \rangle$ (mean $\pm$
  s.d.\ over 24 runs).}
  \label{fig:engagement}
\end{figure*}

Figures \ref{fig:mobility}--\ref{fig:static} establish the assimilative
baselines. In the purely physical model, the competition between the local
convergence time $\tau_{\mathrm{op}} \sim \alpha_{\mathrm{tot}}^{-1}$ and
the neighbourhood-renewal time $\tau_\ell \sim \ell^2/D$ controls how many
opinion clusters survive: $n_c \approx 7 \pm 2$ frozen local clusters for
$D \lesssim 10^{-4}$, merging into the mean-field cluster set
($n_c \approx 2$--$3$ at $\epsilon = 0.3$ \citep{deffuant2000}) for
$D \gtrsim 10^{-2}$, consistent with mobile-agent simulations
\citep{alraddadi2024}. Adding a \emph{static, opinion-blind}
digital layer at low mobility heals this excess fragmentation: a modest
attention share ($\lambda \approx 0.2$--$0.3$) already merges the frozen
clusters down to $n_c \approx 3.3$--$3.5$, close to the mean-field set,
beyond which more digital
attention changes nothing (Fig.~\ref{fig:static}). Opinion-blind
long-range exposure acts, in effect, like infinite-range mobility. (These
levels run at $\sigma_x = 0$; repeating the Level-1 sweep at the adaptive
levels' noise $\sigma_x = 0.02$ reduces the frozen cluster count---noise
merges micro-clusters, $n_c = 4.8 \pm 1.6$ at $D = 10^{-5}$---but leaves
the fragmentation-to-mean-field crossover intact, so the level comparison
is not a noise artefact.) In the
assimilative world, connectivity is benign; everything that follows is
about what curation and repulsion do to this baseline.

\subsection{Algorithmic homophily builds delocalised echo chambers
(Level 3)}
\label{sec:level3}

With similarity-driven rewiring (still purely assimilative influence), the
feedback loop opinion $\to$ topology $\to$ exposure $\to$ opinion closes as
$\gamma$ grows (Fig.~\ref{fig:adaptive}): opinion assortativity climbs to
one, digital modularity rises, and the disagreement visible in an agent's
feed collapses while global fragmentation persists. This is an echo
chamber in the precise sense of \citet{baumann2020}---high global
variance, near-zero visible disagreement---via the mechanism of
\citet{sirbu2019} and \citet{santos2021} transplanted into a coevolving
multiplex \citep{holme2006}. The communities are spatially delocalised:
geography carries no trace of them. In the assimilative world this is the
worst a platform can do in our model: freeze moderate fragmentation and
hide it from its participants.

\subsection{The inversion: uncurated exposure radicalizes (Level 4)}
\label{sec:level4}

Activating the repulsive branch and heavy-tailed influence strengths
(Fig.~\ref{fig:kernels}) produces the inversion
(Fig.~\ref{fig:engagement}). All three designs end with two antagonistic
blocs, but with sharply different degree and speed, in the order predicted
by the two-bloc reduction of Sec.~\ref{sec:twobloc}. The neutral platform
is the most polarizing ($\operatorname{Var}(x) = 0.98 \pm 0.03$ over 24
realisations, opinions
pinned at $\pm 1$ within a few time units): with $p = 1/2$, half of every
feed is repulsion-triggering at all times. The controversy platform is
nearly as extreme ($0.89 \pm 0.03$) but plateaus just short of the boundary
on the simulated horizon ($\langle|x|\rangle \approx 0.95$), as predicted
by the collapse of $p(y)$ when $2y \to 2$. The similarity platform is the least polarizing
($0.66 \pm 0.10$): hiding opposed content starves the repulsive channel,
leaving only the slow creep fed by residual cross-cutting links. Because
the influence strengths have infinite variance, we also checked the
distribution shape: medians ($0.99 / 0.90 / 0.68$) and interquartile
ranges give the same ordering as the means, with no overlap between
platforms. The
network signatures decouple from the opinion signatures: the neutral
platform reaches maximal polarization with \emph{zero} opinion
assortativity ($\rho \approx 0$)---polarization without echo
chambers---whereas both curated platforms end almost perfectly assortative
($\rho \approx 0.9$--$1.0$).

\subsection{Robustness: repulsive-zone exposure orders the designs}
\label{sec:robustness}

\begin{figure*}[!tp]
  \centering
  \includegraphics[width=\linewidth]{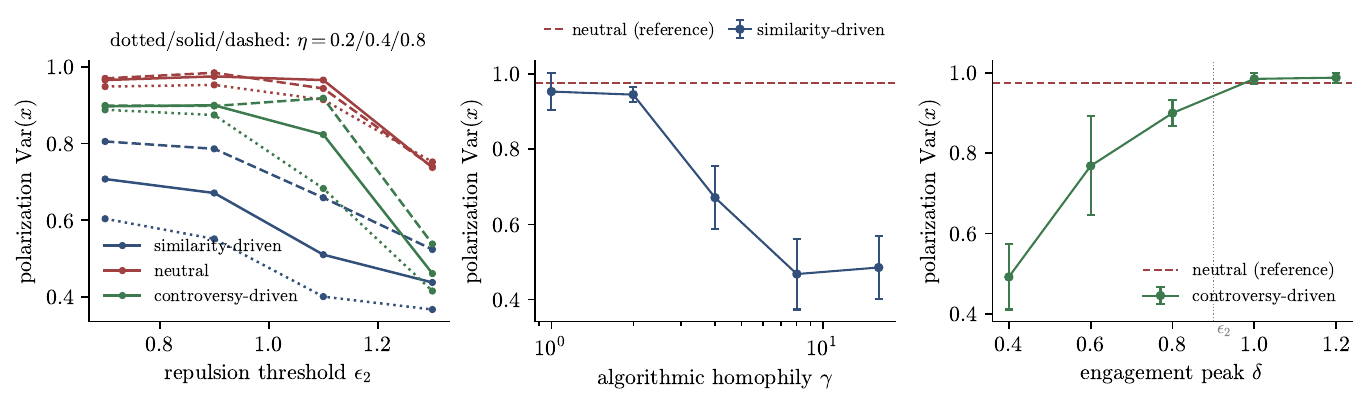}
  \caption{Robustness of the inversion. Left: final polarization versus
  the repulsion threshold $\epsilon_2$ for $\eta = 0.2/0.4/0.8$
  (dotted/solid/dashed); colours denote platforms (blue similarity, red
  neutral, green controversy). Centre: similarity platform versus
  algorithmic homophily $\gamma$, with the neutral platform as reference.
  Right: controversy platform versus engagement peak $\delta$; the dotted
  line marks $\delta = \epsilon_2$. Mean $\pm$ s.d.\ over 6 runs.}
  \label{fig:robustness}
\end{figure*}

\begin{figure}[!tbp]
  \centering
  \includegraphics[width=\linewidth]{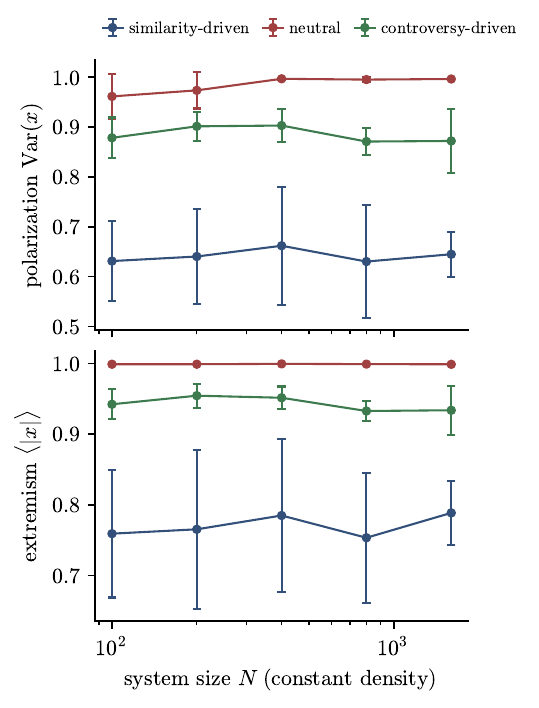}
  \caption{System-size dependence of the inversion at constant density
  ($L = \sqrt{N/200}$), $\lambda = 0.6$. Polarization (top) and extremism
  (bottom) for the three platforms, $N = 100$--$1600$.}
  \label{fig:nscaling}
\end{figure}

\begin{figure*}[!tp]
  \centering
  \includegraphics[width=\linewidth]{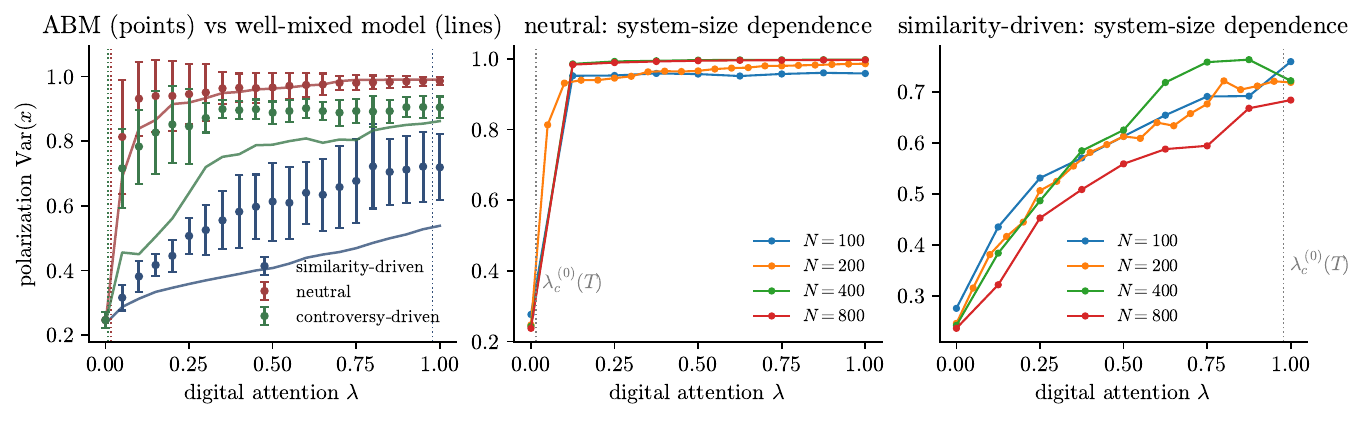}
  \caption{Radicalization onset in the digital attention share. Left:
  final polarization versus $\lambda$ for the three platforms; points are
  the spatial ABM ($N = 200$, 12 seeds), solid lines the well-mixed
  particle model of Eq.~\eqref{eq:mkv} (12 realisations), dotted
  verticals the local-rate crossover estimates $\lambda_c^{(0)}(T)$ of
  Table~\ref{tab:rates}, shown where they fall inside the sampled range
  (no parameters fitted to these data; the exact boundary-reaching
  crossover for the similarity kernel lies outside the physical interval,
  Sec.~\ref{sec:threshold}). Centre and
  right: system-size dependence of the
  $\operatorname{Var}(x)$--$\lambda$ curves for the neutral and similarity
  platforms.}
  \label{fig:threshold}
\end{figure*}

\begin{figure*}[!tp]
  \centering
  \includegraphics[width=\linewidth]{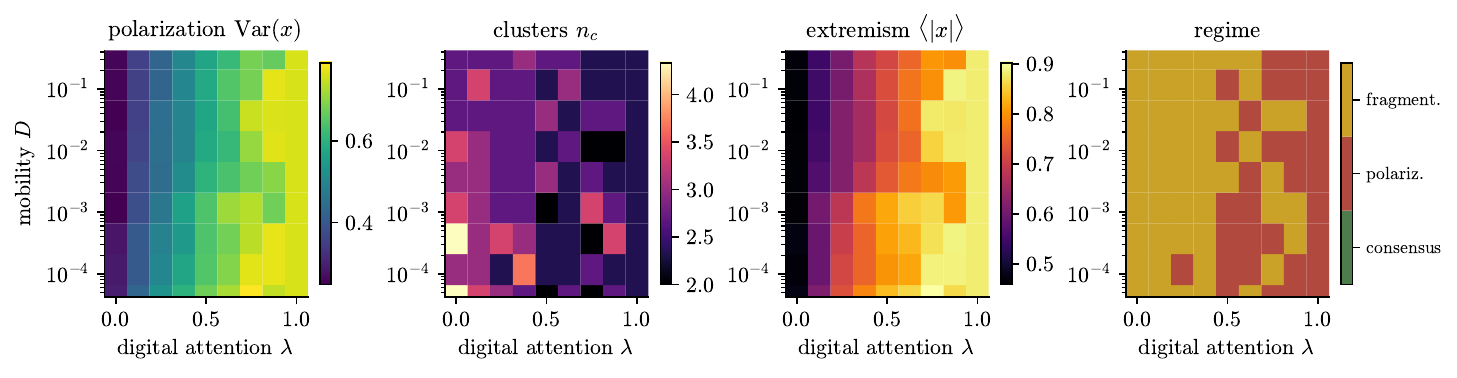}
  \caption{Phase diagram of the full Level-4 model (similarity-driven
  engagement, repulsion, heavy-tailed influence) in the
  mobility--digital-attention plane. From left to right: polarization
  $\operatorname{Var}(x)$, cluster count $n_c$, extremism
  $\langle|x|\rangle$, and the majority categorical state over 3
  realisations per grid point. The panel is a coarse regime map intended
  to reveal qualitative structure; at 3 realisations per point the
  boundaries between regimes are not sharply resolved. At $\lambda = 1$
  the physical layer is off
  and the dynamics is independent of $D$ by construction.}
  \label{fig:phase}
\end{figure*}

Figure~\ref{fig:robustness} tests the ordering across the repulsion and
curation parameters. Across the full $(\eta, \epsilon_2)$ grid the order
neutral $\ge$ controversy $>$ similarity holds in the mean at every grid
point, with separations much larger than the across-seed spread at the
widest repulsive zone ($\epsilon_2 = 0.7$); the gaps narrow as
$\epsilon_2$ grows until the controversy--similarity gap falls within the
seed spread at $\epsilon_2 = 1.3$; all platforms
polarize less as $\epsilon_2$ grows (repulsion becomes harder to trigger)
and more as $\eta$ grows. The two curated platforms interpolate toward the
neutral level in exactly the way the theory predicts. The similarity
platform approaches the neutral bound as $\gamma$ decreases toward zero
(curation too weak to withhold repulsive content)---already at
$\gamma = 1$, the weakest curation simulated, final polarization lies
close to the neutral reference---and its protective effect saturates for
$\gamma \gtrsim 8$. The controversy platform sits below the neutral bound
while its engagement peak $\delta < \epsilon_2$ (much of its exposure
lands in the indifference band) and reaches the bound once
$\delta \ge \epsilon_2$, when engagement maximisation and repulsion
maximisation coincide---at which point it matches the neutral end state
while initially radicalizing at twice the neutral rate
(Table~\ref{tab:rates}). The neutral platform thus realizes the maximal
polarization permitted by the bounded opinion space, and the curated
designs approach this level as their exposure profiles place increasing
weight in the repulsive zone.

Figure~\ref{fig:nscaling} scans the system size at constant density up to
$N = 1600$. The three platforms' polarization and extremism are
essentially flat in $N$: the inversion persists across the tested
system sizes and is not a small-$N$ artefact.

Table~\ref{tab:controls} adds five further controls at the reference
point ($\lambda = 0.6$, 12 realisations each). The inversion survives,
with the same ordering: (i) homogeneous strengths $s_i = 1$---the regime
in which the two-bloc reduction is exact in the mean---so neither the
heavy tail nor influence heterogeneity is needed for the effect; (ii) a
finite-variance strength law ($\kappa = 4$); (iii) reflecting instead of
clipping boundaries, so the saturation ordering is not an artefact of the
projection; (iv) halved and doubled attention windows ($k = 5, 20$),
which shift the similarity platform's creep in the direction predicted by
the slot-fluctuation mechanism (fewer slots, fewer cross-bloc bursts);
and (v) slower and faster rewiring ($\rho = 1, 20$). The last is the most
informative: fast rewiring strengthens homophilic protection
($\operatorname{Var} = 0.54$ at $\rho = 20$) while slow rewiring weakens
it ($0.87$ at $\rho = 1$, where the controversy--similarity gap narrows
into the seed spread)---exactly the dependence expected from the
quasi-stationarity assumption behind Eq.~\eqref{eq:p}, since slowly
renewed feeds let occasional cross-bloc links persist and act for longer.

\begin{table}[!t]
  \caption{Robustness controls at the reference point ($\lambda = 0.6$,
  $T = 80$): final polarization $\operatorname{Var}(x)$, mean $\pm$ s.d.\
  over 12 realisations. The reference row uses 24 realisations.}
  \label{tab:controls}
  \begin{tabular}{lccc}
    \toprule
    Variant & Similarity & Neutral & Controversy \\
    \midrule
    Reference & $0.66 \pm 0.10$ & $0.98 \pm 0.03$ & $0.89 \pm 0.03$ \\
    $s_i = 1$ & $0.68 \pm 0.08$ & $1.00 \pm 0.01$ & $0.91 \pm 0.02$ \\
    $\kappa = 4$ & $0.67 \pm 0.09$ & $0.99 \pm 0.01$ & $0.91 \pm 0.02$ \\
    Reflecting & $0.62 \pm 0.12$ & $0.97 \pm 0.04$ & $0.89 \pm 0.03$ \\
    $k = 5$ & $0.54 \pm 0.08$ & $0.98 \pm 0.02$ & $0.90 \pm 0.03$ \\
    $k = 20$ & $0.76 \pm 0.11$ & $0.98 \pm 0.03$ & $0.92 \pm 0.03$ \\
    $\rho = 1$ & $0.87 \pm 0.08$ & $0.98 \pm 0.02$ & $0.91 \pm 0.03$ \\
    $\rho = 20$ & $0.54 \pm 0.10$ & $0.97 \pm 0.04$ & $0.88 \pm 0.05$ \\
    \bottomrule
  \end{tabular}
\end{table}

\subsection{The radicalization crossover and the well-mixed model}
\label{sec:results-threshold}

\begin{figure*}[!tp]
  \centering
  \includegraphics[width=\linewidth]{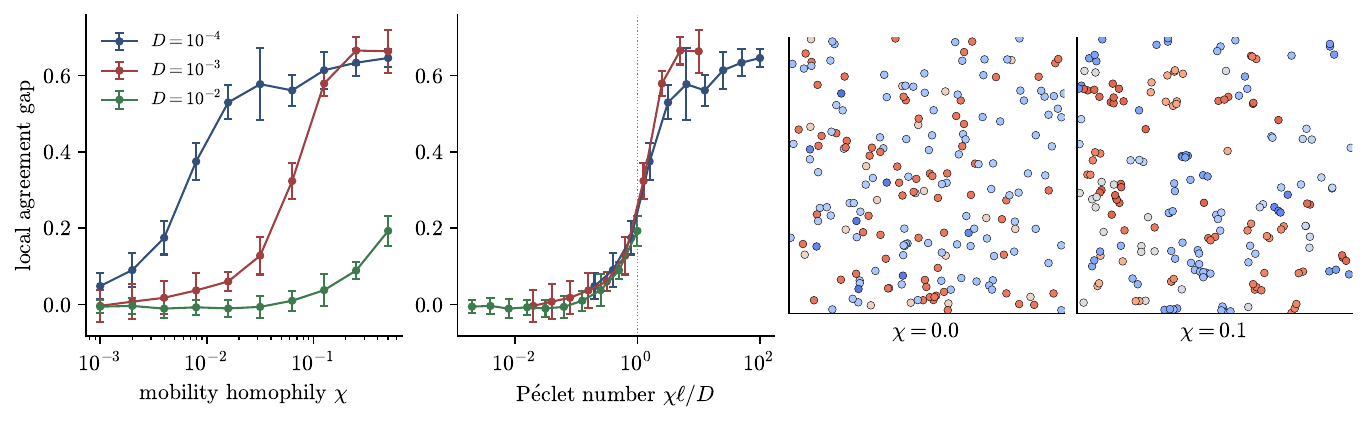}
  \caption{Level 5: homophilic mobility restores geographic opinion
  structure. Left: local agreement gap versus the homophilic drift
  strength $\chi$ for three mobilities. Centre-left: the same data
  against the P\'eclet number $\chi\ell/D$; the curves approximately
  collapse, with onset at $\mathrm{Pe} \sim 1$ (dotted). Right: final
  spatial configurations at $D = 10^{-3}$ without ($\chi = 0$) and with
  ($\chi = 0.1$) homophilic mobility.}
  \label{fig:level5}
\end{figure*}

Figure~\ref{fig:threshold} compares the simulated
$\operatorname{Var}(x)$--$\lambda$ curves with the theory of
Sec.~\ref{sec:theory}. The neutral platform jumps to its saturated value
already at $\lambda = 0.05$--$0.10$, the finest nonzero attention shares
sampled---consistent with its predicted crossover
$\lambda_c \approx 0.016$, which lies below the grid resolution, and
inconsistent with the naive uniform-state threshold ($0.34$). The
controversy platform starts radicalizing equally early
($\lambda_c^{(0)} \approx 0.008$) but rises \emph{gradually}, from
$\operatorname{Var}(x) \approx 0.72$ at $\lambda = 0.05$ to its
$\approx 0.90$ plateau around $\lambda \approx 0.35$--$0.40$---in
agreement with the exact quadrature, which puts the plateau-reaching
crossover at $\lambda_c(80; 0.95) \approx 0.40$
(Sec.~\ref{sec:threshold}), and inconsistent with the naive uniform-state
threshold ($0.55$). Both curves are essentially independent of
system size (centre panel, shown for the neutral platform), as expected
for a finite-horizon crossover
rather than a phase transition. The similarity platform grows smoothly
along $\lambda$ toward a partial radicalization, with no crossover
anywhere in the physical range---as the exact quadrature requires, since
$t_{\mathrm{sim}} \simeq 507.7/\lambda$ exceeds the horizon $T = 80$ even
at $\lambda = 1$; a
weak, non-monotone trend with $N$ (right panel), at the edge of
statistical resolution, is consistent with a partial averaging-out of
slot-level fluctuations---with only $k = 10$ slots, occasional
high-influence cross-bloc sources produce bursts of repulsion that the
population-wide weighting of Eq.~\eqref{eq:mkv} smooths away, which is
also why the well-mixed curve (solid blue) sits below the ABM points
while tracking their shape. The well-mixed particle model (12
realisations) reproduces the neutral curve nearly quantitatively for
$\lambda \gtrsim 0.2$ (within $\approx 0.03$, and within $\approx 0.01$
for $\lambda \gtrsim 0.45$; at the steep onset $\lambda = 0.05$--$0.15$
it lags by up to $\approx 0.12$) and sits below the controversy curve at
all $\lambda$---by up to $\approx 0.33$ near the onset, narrowing to
$\approx 0.04$ at $\lambda = 1$, for the same slot-fluctuation
reason---with no parameters fitted to these data.

\subsection{Phase diagram: mobility dependence vanishes at high digital
attention}
\label{sec:phase}

Figure~\ref{fig:phase} assembles the global picture in the $(D, \lambda)$
plane, computed for the similarity-driven design---the slowest to
radicalize, so the boundaries below are conservative; the other designs
saturate at far lower $\lambda$. At $\lambda \approx 0$ the physical
layer decides the outcome and
the Level-1 crossover is recovered: frozen local fragmentation at low $D$,
the mean-field cluster set at high $D$. As $\lambda$ grows the platform
captures the attention budget, the repulsive channel activates through the
residual cross-cutting exposure, and both polarization and extremism rise
across \emph{all} mobilities: the high-$\lambda$ region is a bimodal,
radicalised state essentially independent of $D$. Once the platform
curates most of an agent's exposure, it no longer matters how fast people
move: geography sets the outcome only for societies whose attention it
still owns.

\subsection{Geography: a null result and its repair (Levels 1 and 5)}
\label{sec:geography}

The bottom panel of Fig.~\ref{fig:mobilitysweep} contains a null result
that qualifies a common intuition. Moran's $I$---and the local agreement
gap---never departs from noise level, \emph{even in the frozen regime
where interactions are strictly local}. The first-order part of this
statement is a symmetry: on the torus, the initial law and the dynamics
with $\chi = 0$ are translation invariant, so
$\mathbb{E}[x_i \mid \rr_i = \rr]$ is independent of $\rr$ at all
times---no deterministic opinion map can exist. Symmetry does not forbid
\emph{pair}-level structure (nearby agents interacted more), but the
measured pair correlations also stay at noise level: the ensemble mean of
Moran's $I$ is compatible with zero at every $D$ (largest excursion
$2.2$ standard errors among eleven mobilities at $6$ seeds each, so the
null would not detect a weak true correlation), with an across-seed
spread of $\approx 0.13$, and the local agreement gap remains at
$0.018 \pm 0.027$ even in the frozen regime---an order of magnitude below
the $0.4$--$0.7$ signal that develops once opinion--position coupling is
switched on (Fig.~\ref{fig:level5}). Disagreeing agents
interleave at the same locations, interacting with disjoint compatible
subsets, because nothing couples opinion to position. Locality controls
\emph{how many} opinions survive but not \emph{where} they live. Within
this model, low mobility alone is therefore insufficient to generate
geographic echo chambers; persistent spatial domains require
opinion--position coupling strong enough to compete with diffusion.

Level 5 quantifies how much coupling is needed. With the homophilic drift
of Eq.~\eqref{eq:motion} ($\chi > 0$: approach compatible neighbours,
avoid incompatible ones \citep{pasimeni2025, djurdjevac2024}), spatial
opinion domains appear (Fig.~\ref{fig:level5}). The control parameter is
the P\'eclet number comparing drift and diffusion over the interaction
range,
\begin{equation}
  \mathrm{Pe} = \frac{\chi\,\ell}{D},
  \label{eq:peclet}
\end{equation}
and the local agreement gap for different mobilities approximately
collapses onto a single curve in $\mathrm{Pe}$, rising from near noise
level for $\mathrm{Pe} \lesssim 0.3$ to strong spatial segregation for
$\mathrm{Pe} \gtrsim 3$. Geographic opinion structure therefore exhibits a P\'eclet-controlled
crossover in the strength of opinion-dependent mobility: it is weak below
$\mathrm{Pe}\lesssim0.3$ and pronounced above $\mathrm{Pe}\gtrsim3$ in
the explored range. The Brownian model of Levels 1--4 has $\chi=0$ and
therefore provides the natural null baseline by construction.

\section{Discussion}
\label{sec:discussion}

\subsection{Summary} We combined bounded-confidence opinion dynamics,
Brownian and homophilic mobility, and an algorithmically curated digital
layer under a finite attention budget, and found that the interaction
between platform design and influence psychology---not either
alone---decides the collective outcome. With assimilative influence, the
neutral platform is a defragmenting force and homophilic curation builds
echo chambers; with the repulsive channel active, all three tested
designs exhibit outward radicalizing drift, but with strongly different
finite-horizon magnitudes. At the reference Level-4 attention share the
neutral platform reaches the maximal polarization permitted by the bounded
opinion space, controversy-driven curation approaches a strong plateau,
and homophilic curation delays the outward drift. The two-bloc reduction,
the exact radicalization quadrature and its finite-horizon crossover, and
the well-mixed comparison make the mechanism analytical rather than
merely observational; the robustness and finite-size scans show that the
inversion persists across the tested parameter and system-size ranges. A methodological implication follows: the na\"ive
linear-stability analysis of the uniform state predicts thresholds an
order of magnitude too large, because fragmentation preempts it---in
coupled opinion--network systems, the state that matters for stability is
the one the fast dynamics builds first.

\subsection{Relation to experiments} The inversion offers a candidate
population-scale mechanism consistent with the field experiment of
\citet{bail2018}, in which a month of bot-curated exposure to opposing
views made Republican participants substantially more conservative (the
corresponding shift for Democrats was not statistically significant---an
asymmetry that our symmetric, one-dimensional model does not address): in
our terms, an intervention that raises cross-cutting exposure moves a
similarity-curated population toward the neutral level, and when the
repulsive channel dominates, that movement increases polarization. We
stress that this comparison is qualitative---no model parameter is
calibrated to data---and that the repulsive channel itself is empirically
contested (Sec.~\ref{sec:model}): the supporting longitudinal evidence
\citep{tang2025} concerns a single non-political preference item
(adolescents' taste in clothing style), finds little support for bounded
confidence itself, and implies only low macro-level polarization in its
own calibrated simulations, while controlled experiments find little
evidence of negative shifts at all \citep{takacs2016}. The
field-experimental record on exposure interventions is equally mixed:
counter-attitudinal exposure has also \emph{reduced} affective
polarization \citep{levy2021}, and large platform experiments altering
feed composition found little attitudinal effect \citep{nyhan2023}. Read
this way, the model turns a psychological question into a measurable
platform-policy question: it predicts intervention effects of
\emph{either} sign depending on the prevalence of negative influence in
the target population---consistent with, though certainly not established
by, the heterogeneous field record. When influence is assimilative, added
exposure diversity heals fragmentation (Sec.~\ref{sec:baselines});
interventions on recommendation algorithms that ignore which influence
regime the population is in can therefore backfire in either direction.
At the collective level, our neutral-platform result reaches, by a
different microscopic route---negative influence on a single opinion
dimension rather than partisan sorting across many identity
dimensions---a related qualitative conclusion to \citet{tornberg2022}: digital
interaction can increase polarization through exposure beyond local or
like-minded social neighborhoods rather than through isolation alone. Recent voter-type models with
involvement likewise find polarization rising with connectivity itself
\citep{pham2026}; our mechanism differs by identifying repulsive
cross-bloc exposure under a conserved attention budget as the kinetic
control parameter.

\subsection{The geography null result} Under opinion-independent mobility
we detect no geographic opinion structure to destroy, within the
resolution of our simulations: position and opinion stay uncorrelated at
every mobility, by symmetry at first order and at pair level within the
stated statistical power. Within this model, robust geographic opinion
structure therefore requires an opinion--position coupling---%
homophilic migration, socially structured mobility, or spatially
correlated exogenous influence---and Level 5 shows the coupling need only
cross a P\'eclet-scale crossover centred around $\chi\ell/D \sim 1$.
This sharpens,
rather than contradicts, the mobile-agent literature
\citep{alraddadi2024, pasimeni2025}: what matters is not whether agents
move, but whether their movement knows about opinions.

\subsection{Limitations and outlook} Opinions are one-dimensional and
confidence bounds static; adaptive-confidence extensions are natural
\citep{li2025}. Opinions also live on a bounded interval, so
``radicalization'' here means pinning at the boundary of opinion space;
in unbounded formulations \citep{baumann2020} the analogue is unbounded
divergence. The inversion mechanism---the cross-bloc exposure rate
$p(y)$---does not depend on this choice, but the saturation values
$\operatorname{Var}(x) \approx 1$ do. The two-bloc reduction is heuristic---it assumes
symmetric blocs and fast rewiring---and its quantitative success invites
a rigorous analysis of the well-mixed dynamics \eqref{eq:mkv}---its
McKean--Vlasov limit for $\kappa > 3$, the non-self-averaging regime at
heavier tails, and the metastability structure behind the finite-horizon
crossover \citep{castellano2009, bernardo2024}. The attention budget is a fixed
split rather than a dynamic allocation; letting $\lambda$ itself be
optimised by an engagement-maximising platform closes an obvious loop.
Stubborn agents are implemented but unexplored; sweeping their number and
placement against platform designs connects to opinion control
\citep{tian2018}. The model also equates the engagement kernel with what
agents attend to \emph{and are influenced by}; real engagement
optimisation responds to behaviour rather than opinion distance, and
exposure need not translate into influence. The quantitative platform ranking is therefore a finite-time statement
at the explored parameters and horizons, not a universal asymptotic order.
Within the deterministic two-bloc reduction, any design with $p(y)>0$
throughout the traversed range has the same boundary-pinned asymptotic
destination, although the associated timescale can become arbitrarily
long. Moreover, when two exposure profiles cross---as the controversy
profile does with the neutral baseline---their relative ranking depends on
the chosen target and observation horizon through Eq.~\eqref{eq:tradgen}. Finally, the
model's sharpest empirical claim---that
exposure diversity helps or harms depending on the prevalence of negative
influence---is testable in principle by conditioning field-experiment
outcomes on measured latitudes of acceptance.

\section*{Data and code availability}
The full simulation code, the scripts that generate every figure and table,
the raw sweep data, and animations of the coupled dynamics are openly
available at \url{https://github.com/REsteche/social-modeling}.

\section*{Funding}
This research did not receive any specific grant from funding agencies in
the public, commercial, or not-for-profit sectors.

\section*{Declaration of generative AI and AI-assisted technologies in the manuscript preparation process}

During the preparation of this work, the author(s) used ChatGPT Deep Research tooling for reference gathering 
and coding assistance for the simulations. The author(s) reviewed and edited the output as needed and take 
full responsibility for the content of the published article. 

\bibliographystyle{elsarticle-num-names}
\bibliography{references}

@article{deffuant2000,
  author  = {Deffuant, Guillaume and Neau, David and Amblard, Fr{\'e}d{\'e}ric and Weisbuch, G{\'e}rard},
  title   = {Mixing beliefs among interacting agents},
  journal = {Advances in Complex Systems},
  volume  = {3},
  number  = {01n04},
  pages   = {87--98},
  year    = {2000},
  doi     = {10.1142/S0219525900000078}
}

@article{hegselmann2002,
  author  = {Hegselmann, Rainer and Krause, Ulrich},
  title   = {Opinion dynamics and bounded confidence: models, analysis and simulation},
  journal = {Journal of Artificial Societies and Social Simulation},
  volume  = {5},
  number  = {3},
  year    = {2002},
  url     = {https://www.jasss.org/5/3/2.html}
}

@article{lorenz2007,
  author  = {Lorenz, Jan},
  title   = {Continuous opinion dynamics under bounded confidence: a survey},
  journal = {International Journal of Modern Physics C},
  volume  = {18},
  number  = {12},
  pages   = {1819--1838},
  year    = {2007},
  doi     = {10.1142/S0129183107011789}
}

@article{castellano2009,
  author  = {Castellano, Claudio and Fortunato, Santo and Loreto, Vittorio},
  title   = {Statistical physics of social dynamics},
  journal = {Reviews of Modern Physics},
  volume  = {81},
  number  = {2},
  pages   = {591--646},
  year    = {2009},
  doi     = {10.1103/RevModPhys.81.591}
}

@article{bernardo2024,
  author  = {Bernardo, Carmela and Altafini, Claudio and Proskurnikov, Anton and Vasca, Francesco},
  title   = {Bounded confidence opinion dynamics: A survey},
  journal = {Automatica},
  volume  = {159},
  pages   = {111302},
  year    = {2024},
  doi     = {10.1016/j.automatica.2023.111302}
}

@article{alraddadi2024,
  author  = {Alraddadi, Enas E. and Allen, Stuart M. and Colombo, Gualtiero B. and Whitaker, Roger M.},
  title   = {A novel framework to classify opinion dynamics of mobile agents under the bounded confidence model},
  journal = {Adaptive Behavior},
  volume  = {32},
  number  = {2},
  pages   = {167--187},
  year    = {2024},
  doi     = {10.1177/10597123231195423}
}

@article{holme2006,
  author  = {Holme, Petter and Newman, Mark E. J.},
  title   = {Nonequilibrium phase transition in the coevolution of networks and opinions},
  journal = {Physical Review E},
  volume  = {74},
  number  = {5},
  pages   = {056108},
  year    = {2006},
  doi     = {10.1103/PhysRevE.74.056108}
}

@article{antonopoulos2018,
  author  = {Antonopoulos, Chris G. and Shang, Yilun},
  title   = {Opinion formation in multiplex networks with general initial distributions},
  journal = {Scientific Reports},
  volume  = {8},
  pages   = {2852},
  year    = {2018},
  doi     = {10.1038/s41598-018-21054-0}
}

@article{baumann2020,
  author  = {Baumann, Fabian and Lorenz-Spreen, Philipp and Sokolov, Igor M. and Starnini, Michele},
  title   = {Modeling echo chambers and polarization dynamics in social networks},
  journal = {Physical Review Letters},
  volume  = {124},
  number  = {4},
  pages   = {048301},
  year    = {2020},
  doi     = {10.1103/PhysRevLett.124.048301}
}

@article{sirbu2019,
  author  = {S{\^i}rbu, Alina and Pedreschi, Dino and Giannotti, Fosca and Kert{\'e}sz, J{\'a}nos},
  title   = {Algorithmic bias amplifies opinion fragmentation and polarization: A bounded confidence model},
  journal = {PLOS ONE},
  volume  = {14},
  number  = {3},
  pages   = {e0213246},
  year    = {2019},
  doi     = {10.1371/journal.pone.0213246}
}

@article{santos2021,
  author  = {Santos, Fernando P. and Lelkes, Yphtach and Levin, Simon A.},
  title   = {Link recommendation algorithms and dynamics of polarization in online social networks},
  journal = {Proceedings of the National Academy of Sciences},
  volume  = {118},
  number  = {50},
  pages   = {e2102141118},
  year    = {2021},
  doi     = {10.1073/pnas.2102141118}
}

@article{dearruda2022,
  author  = {de Arruda, Henrique Ferraz and Cardoso, Felipe Maciel and de Arruda, Guilherme Ferraz and Hern{\'a}ndez, Alexis R. and Costa, Luciano da Fontoura and Moreno, Yamir},
  title   = {Modelling how social network algorithms can influence opinion polarization},
  journal = {Information Sciences},
  volume  = {588},
  pages   = {265--278},
  year    = {2022},
  doi     = {10.1016/j.ins.2021.12.069}
}

@article{kurmyshev2011,
  author  = {Kurmyshev, Evguenii and Ju{\'a}rez, H{\'e}ctor A. and Gonz{\'a}lez-Silva, Ricardo A.},
  title   = {Dynamics of bounded confidence opinion in heterogeneous social networks: Concord against partial antagonism},
  journal = {Physica A: Statistical Mechanics and its Applications},
  volume  = {390},
  number  = {16},
  pages   = {2945--2955},
  year    = {2011},
  doi     = {10.1016/j.physa.2011.03.037}
}

@article{tang2025,
  author  = {Tang, Tanzhe and Snijders, Tom A. B. and Flache, Andreas},
  title   = {An empirical and simulation investigation of bounded confidence and negative influence in opinion dynamics using stochastic actor-oriented modelling},
  journal = {Journal of Artificial Societies and Social Simulation},
  volume  = {28},
  number  = {1},
  pages   = {2},
  year    = {2025},
  doi     = {10.18564/jasss.5566}
}

@article{pal2023,
  author  = {Pal, Ritam and Kumar, Aanjaneya and Santhanam, M. S.},
  title   = {Depolarization of opinions on social networks through random nudges},
  journal = {Physical Review E},
  volume  = {108},
  number  = {3},
  pages   = {034307},
  year    = {2023},
  doi     = {10.1103/PhysRevE.108.034307}
}

@misc{starnini2025,
  author       = {Starnini, Michele and Baumann, Fabian and Galla, Tobias and Garc{\'i}a, David and {\'I}{\~n}iguez, Gerardo and Karsai, M{\'a}rton and Lorenz, Jan and Sznajd-Weron, Katarzyna},
  title        = {Opinion dynamics: Statistical physics and beyond},
  year         = {2025},
  eprint       = {2507.11521},
  archivePrefix = {arXiv},
  note         = {Accepted in Reviews of Modern Physics (2026)},
  doi          = {10.1103/j1zg-ddqv}
}

@article{pansanella2022,
  author  = {Pansanella, Valentina and Rossetti, Giulio and Milli, Letizia},
  title   = {Modeling algorithmic bias: simplicial complexes and evolving network topologies},
  journal = {Applied Network Science},
  volume  = {7},
  pages   = {57},
  year    = {2022},
  doi     = {10.1007/s41109-022-00495-7}
}

@article{axelrod2021,
  author  = {Axelrod, Robert and Daymude, Joshua J. and Forrest, Stephanie},
  title   = {Preventing extreme polarization of political attitudes},
  journal = {Proceedings of the National Academy of Sciences},
  volume  = {118},
  number  = {50},
  pages   = {e2102139118},
  year    = {2021},
  doi     = {10.1073/pnas.2102139118}
}

@article{sabinmiller2020,
  author  = {Sabin-Miller, David and Abrams, Daniel M.},
  title   = {When pull turns to shove: A continuous-time model for opinion dynamics},
  journal = {Physical Review Research},
  volume  = {2},
  number  = {4},
  pages   = {043001},
  year    = {2020},
  doi     = {10.1103/PhysRevResearch.2.043001}
}

@article{bellina2023,
  author  = {Bellina, Alessandro and Castellano, Claudio and Pineau, Paul and Iannelli, Giulio and De Marzo, Giordano},
  title   = {Effect of collaborative-filtering-based recommendation algorithms on opinion polarization},
  journal = {Physical Review E},
  volume  = {108},
  number  = {5},
  pages   = {054304},
  year    = {2023},
  doi     = {10.1103/PhysRevE.108.054304}
}

@inproceedings{cinus2022,
  author    = {Cinus, Federico and Minici, Marco and Monti, Corrado and Bonchi, Francesco},
  title     = {The effect of people recommenders on echo chambers and polarization},
  booktitle = {Proceedings of the Sixteenth International AAAI Conference on Web and Social Media},
  volume    = {16},
  pages     = {90--101},
  year      = {2022},
  doi       = {10.1609/icwsm.v16i1.19275}
}

@article{flache2017,
  author  = {Flache, Andreas and M{\"a}s, Michael and Feliciani, Thomas and Chattoe-Brown, Edmund and Deffuant, Guillaume and Huet, Sylvie and Lorenz, Jan},
  title   = {Models of social influence: Towards the next frontiers},
  journal = {Journal of Artificial Societies and Social Simulation},
  volume  = {20},
  number  = {4},
  pages   = {2},
  year    = {2017},
  doi     = {10.18564/jasss.3521}
}

@article{levy2021,
  author  = {Levy, Ro'ee},
  title   = {Social media, news consumption, and polarization: Evidence from a field experiment},
  journal = {American Economic Review},
  volume  = {111},
  number  = {3},
  pages   = {831--870},
  year    = {2021},
  doi     = {10.1257/aer.20191777}
}

@article{nyhan2023,
  author  = {Nyhan, Brendan and Settle, Jaime and Thorson, Emily and Wojcieszak, Magdalena and Barber{\'a}, Pablo and others},
  title   = {Like-minded sources on {Facebook} are prevalent but not polarizing},
  journal = {Nature},
  volume  = {620},
  pages   = {137--144},
  year    = {2023},
  doi     = {10.1038/s41586-023-06297-w}
}

@article{tian2018,
  author  = {Tian, Ye and Wang, Long},
  title   = {Opinion dynamics in social networks with stubborn agents: An issue-based perspective},
  journal = {Automatica},
  volume  = {96},
  pages   = {213--223},
  year    = {2018},
  doi     = {10.1016/j.automatica.2018.06.041}
}

@article{jager2005,
  author  = {Jager, Wander and Amblard, Fr{\'e}d{\'e}ric},
  title   = {Uniformity, bipolarization and pluriformity captured as generic stylized behavior with an agent-based simulation model of attitude change},
  journal = {Computational and Mathematical Organization Theory},
  volume  = {10},
  number  = {4},
  pages   = {295--303},
  year    = {2005}
}

@article{takacs2016,
  author  = {Tak{\'a}cs, K{\'a}roly and Flache, Andreas and M{\"a}s, Michael},
  title   = {Discrepancy and disliking do not induce negative opinion shifts},
  journal = {PLOS ONE},
  volume  = {11},
  number  = {6},
  pages   = {e0157948},
  year    = {2016},
  doi     = {10.1371/journal.pone.0157948}
}

@article{tornberg2022,
  author  = {T{\"o}rnberg, Petter},
  title   = {How digital media drive affective polarization through partisan sorting},
  journal = {Proceedings of the National Academy of Sciences},
  volume  = {119},
  number  = {42},
  pages   = {e2207159119},
  year    = {2022},
  doi     = {10.1073/pnas.2207159119}
}

@article{bail2018,
  author  = {Bail, Christopher A. and Argyle, Lisa P. and Brown, Taylor W. and Bumpus, John P. and Chen, Haohan and Hunzaker, M. B. Fallin and Lee, Jaemin and Mann, Marcus and Merhout, Friedolin and Volfovsky, Alexander},
  title   = {Exposure to opposing views on social media can increase political polarization},
  journal = {Proceedings of the National Academy of Sciences},
  volume  = {115},
  number  = {37},
  pages   = {9216--9221},
  year    = {2018},
  doi     = {10.1073/pnas.1804840115}
}

@article{djurdjevac2024,
  author  = {Djurdjevac Conrad, Nata{\v{s}}a and Vu, Nhu Quang and Nagel, S{\"o}ren},
  title   = {Co-evolving networks for opinion and social dynamics in agent-based models},
  journal = {Chaos},
  volume  = {34},
  number  = {9},
  pages   = {093116},
  year    = {2024},
  doi     = {10.1063/5.0226054}
}

@article{pasimeni2025,
  author  = {Pasimeni, Francesco and Wade, Richard and Alkemade, Floortje},
  title   = {Opinion dynamic and social clustering in a {2D} space: An agent based experiment},
  journal = {Computational Economics},
  volume  = {67},
  pages   = {3449--3484},
  year    = {2026},
  doi     = {10.1007/s10614-025-10961-w}
}

@article{li2025,
  author  = {Li, Grace J. and Luo, Jiajie and Porter, Mason A.},
  title   = {Bounded-confidence models of opinion dynamics with adaptive confidence bounds},
  journal = {SIAM Journal on Applied Dynamical Systems},
  volume  = {24},
  number  = {2},
  pages   = {994--1041},
  year    = {2025},
  doi     = {10.1137/23M1558951}
}

@article{pham2026,
  author  = {Pham, Tuan Minh and Redner, Sidney and Waldorp, Lourens and Armas, Jay and van der Maas, Han L. J.},
  title   = {Polarization in increasingly connected societies},
  journal = {Physical Review E},
  volume  = {113},
  number  = {5},
  pages   = {054303},
  year    = {2026},
  doi     = {10.1103/5lwq-2n8m}
}

@book{resnick2007,
  author    = {Resnick, Sidney I.},
  title     = {Heavy-Tail Phenomena: Probabilistic and Statistical Modeling},
  publisher = {Springer},
  address   = {New York},
  year      = {2007},
  doi       = {10.1007/978-0-387-45024-7}
}

\end{document}